\documentclass[10pt,journal=jctcce,manuscript=article,layout=twocolumn]{achemso}
\usepackage{iftex}
\usepackage{amsmath,amssymb,amsfonts} 
\usepackage[version=4]{mhchem}
\usepackage[super]{nth}
\usepackage{xfrac,calc,multirow}
\usepackage{nameref,doi,lineno}
\usepackage{color,graphicx}

\ifXeTeX%
\usepackage{mathspec}
\setallmainfonts{Minion Pro}
\usepackage{polyglossia}
\setmainlanguage{english}
\else
\usepackage{mathptmx}
\fi

\setkeys{acs}{articletitle=true}
\setkeys{acs}{doi=true}

\usepackage[noabbrev,capitalize]{cleveref}
\crefname{equation}{eq~}{eqs~}     % abbreviation without a dot
\creflabelformat{equation}{#2#1#3} % no parenthesis
\newcommand{\crefx}[2]{\cref{#1}\,(#2)}
\newcommand{\sifigref}[1]{\figurename\:S{#1}}
\newcommand{\sitabref}[1]{\tablename\:S{#1}}

\graphicspath{{figs/}}
\newcommand{\figwidth}{\ifdim\linewidth>4in .6667\else 1\fi} % reasonable size in a single column mode

\newcommand\Tstrut{\rule{0pt}{2.6ex}}       % top strut for tables with superscripts
\newcommand\Bstrut{\rule[-1.2ex]{0pt}{0pt}} % bottom strut for tables with subscripts
\ifXeTeX%
\else%
\fi%

\usepackage[separate-uncertainty=false]{siunitx}
\DeclareSIUnit\molar{\mole\per\cubic\deci\metre}
\DeclareSIUnit\Molar{\textsc{m}}

\newcommand*{\diff}{\mathop{\mathrm{d}\!}}

\makeatletter
\@ifclasswith{achemso}{layout=twocolumn}
{
	\newcommand{\change}[1]{#1}
	\newcommand{\enlargecolumn}[1]{\enlargethispage{#1}}
}
{
	\newcommand{\change}[1]{{\color{blue}#1}}
	\newcommand{\enlargecolumn}[1]{\relax}
	\linenumbers
}
\makeatother

\hypersetup{pdftitle={A temperature-dependent implicit-solvent model of polyethylene glycol in aqueous solution}, pdfauthor={Richard Chudoba, Jan Heyda, Joachim Dzubiella}}

\title{A temperature-dependent implicit-solvent model of polyethylene glycol in aqueous solution}

\author{Richard Chudoba}
\affiliation{Institut f\"ur Physik, Humboldt-Universit\"at zu Berlin, Newtonstr.~15, D-12489 Berlin, Germany}
\alsoaffiliation{Institut f\"ur Weiche Materie und Funktionale Materialen, Helmholtz-Zentrum Berlin, Hahn-Meitner Platz 1, D-14109 Berlin, Germany}
\email{richard.chudoba@helmholtz-berlin.de}
\author{Jan Heyda}
\affiliation{Department of Physical Chemistry, University of Chemistry and Technology, Technick\'a 5, CZ-16628 Praha 6, Czech Republic}
\email{jan.heyda@vscht.cz}
\author{Joachim Dzubiella}
\affiliation{Institut f\"ur Physik, Humboldt-Universit\"at zu Berlin, Newtonstr.~15, D-12489 Berlin, Germany}
\alsoaffiliation{Institut f\"ur Weiche Materie und Funktionale Materialen, Helmholtz-Zentrum Berlin, Hahn-Meitner Platz 1, D-14109 Berlin, Germany}
\email{joachim.dzubiella@helmholtz-berlin.de}

\begin{tocentry}
	\centering \includegraphics{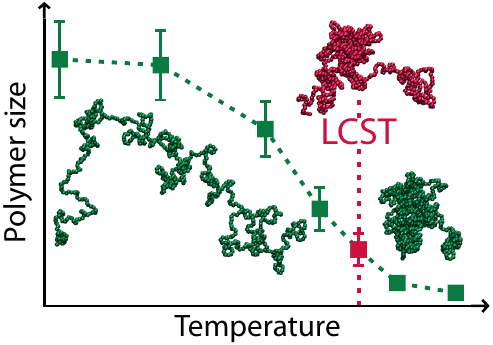}
\end{tocentry}

\begin{document}	
	\begin{abstract}
		A temperature ($T$)-dependent coarse-grained (CG) Hamiltonian of polyethylene glycol/oxide (PEG/PEO) in aqueous solution is reported to be used in implicit-solvent material models in a wide temperature (i.e., solvent quality) range. The $T$-dependent nonbonded CG interactions are derived from a combined “bottom-up” and “top-down” approach. The pair potentials calculated from atomistic replica-exchange molecular dynamics simulations in combination with the iterative Boltzmann inversion are post-refined by benchmarking to experimental data of the radius of gyration. For better handling and a fully continuous transferability in $T$-space, the pair potentials are conveniently truncated and mapped to an analytic formula with three structural parameters expressed as explicit continuous functions of $T$. It is then demonstrated that this model without further adjustments successfully reproduces other experimentally known key thermodynamic properties of semi-dilute PEG solutions such as the full equation of state (i.e., $T$-dependent osmotic pressure) for various chain lengths as well as their cloud point (or collapse) temperature.
	\end{abstract}
	
	\maketitle
	
	\section{Introduction}
	\enlargecolumn{5pt}
	One of the chemically simplest but increasingly popular polymer in material science is polyethylene glycol (PEG), also known as polyethylene oxide (PEO).  
	For instance, PEG is frequently used in biochemistry and biophysics mostly due to its bioinert properties, e.g., as antifouling agent, covalent modifier, carrier matrix, or crowding agent for protein stabilization or crystallization,\cite{Wei2014, McPherson1976,Galkin2000, Knowles2011} just to name a few. Another line of important modern applications of PEG concerns thermo- and stimuli-responsive materials which under the action of external stimuli, most prominently a temperature change, undergo rapid and reversible changes of some of their properties in aqueous environment.\cite{Stuart2010, Ward2011, Gibson2013, Phillips2015, Gandhi2015} PEG itself is thermoresponsive, however, has a rather high lower critical solution temperature (LCST) at $\approx\SI{373}{\kelvin}$ in pure water.\cite{Saeki1976, Kjellander1981} Nevertheless its purposeful integration in aqueous-based materials at more relevant operating conditions can be achieved via copolymerization or by adding cosolvents, leading to complex block-copolymer architectures in solvent mixtures acting as soft switchable and functional materials.\cite{Stuart2010, Deyerle2011}
	
	In order to rationalize and guide future PEG-based soft material design, the quantitative understanding and theoretical description of PEG properties starting from the microscopic structure in aqueous solution to consistent macroscopic observables is a precondition. Due to the chemical simplicity and relatively small sizes of the monomer, atomistic level computer simulations are feasible, at least for small molecular weights.\cite{Lee2008, Starovoytov2011, Fuchs2012, Hezaveh2012} Such simulations allow to obtain details about the polymer conformations,\cite{Tasaki1996, Smith2000} mean polymer size,\cite{Oh2013} or hydration properties\cite{Liese2016} in aqueous solutions, but are restricted to rather short polymer chains and sub-microsecond timescales. Larger scale simulations (longer or more chains) for the modeling of larger assemblies of material components are not feasible on the atomistic level, and efficient mesoscale polymer models that are transferable between various system conditions are in urgent need.\cite{Carbone2008, Qian2008, Krishna2009, Abbott2015, DeSilva2017}
	
	To bridge the scales, several coarse-grained (CG) models of PEG have indeed been proposed with varying degree of chemical detail. 
	Most CG approaches start from explicit-solvent atomistic simulations, so called “bottom-up” approaches, and thus capture faithfully the essential polymer chemical features and local structure. The coarse-graining is achieved by established statistical mechanics methods that integrate out the microscopic degrees of freedom,  such as the iterative Boltzmann inversion (IBI).\cite{Schommers1983, Reith2003, Rosenberger2016} In this or similar ways, fully implicit-solvent PEG coarse-grained models were derived\cite{Bedrov2006, Fischer2008a, Cordeiro2010} as well as models where water molecules are coarse-grained to single neutral beads.\cite{Prasitnok2013}  Other approaches have rather followed a “top-down” strategy by resigning on chemical details or solvent and using simple empirical pair interaction potentials between monomers, such as Lennard-Jones or Weeks-Chandler-Anderson,\cite{Weeks1971} which aim to reproduce a variety of known experimental observables, such as the polymer size or the osmotic pressure.\cite{Jeppesen1996, Xie2016} \change{Indeed, under good solvent conditions and in the limit of infinitely long PEG chains, it was found that available macroscopic experimental data can be reproduced.\cite{Xie2016}}
	Finally, several mesoscale models for PEG have combined bottom-up and top-down approaches to obtain consistent descriptions of multiple structural and macroscopic properties at the same time,\cite{Shinoda2007, Shinoda2008, Jusufi2011} including those based on the MARTINI-approach.\cite{Lee2009, Choi2014, Wang2015, Rossi2012, Nawaz2014, Taddese2017}
	
	Despite the substantial body of mesoscale models of aqueous PEG published up to now, without exception they were parametrized close to room temperature ($\approx\SI{298}{\kelvin}$), thus focusing only on the well soluble state, that is, in a good solvent state. Hence, a large gap exists within the model range concerning the models’ transferability to other temperatures, in particular regarding configurational and miscibility properties of PEG for increasing temperatures moving towards the $\theta$ and critical solution temperatures. Recall that aqueous PEG solutions possess an LCST (opposite to an upper CST) and thus are macromolecular systems whose \emph{effective} polymer-polymer attraction is governed by positive entropy. This is opposite to the simple classical interaction Hamiltonians that are commonly used in textbook modeling,\cite{Flory1953} and are still applicable for polymer melts,\cite{Carbone2008, Qian2008, Krishna2009} where the intricate solvent effects play no role.
	For PEG in water, the nonbonded self-interactions of the polymer become more attractive for increasing temperature, i.e., the solvent eventually turns bad, driving a chain collapse transition (for a single chain\cite{Wu1998}) and clouding (for many chains) at the LCST when approaching from lower temperatures. Hence, even for small temperature changes close to room temperature it is  \latin{a~priori} unclear to which extent the previously introduced models, which define purely energetic nonbonded potentials, can be applied and if they perform at least qualitatively correct. Moreover, it is questionable, if and how they can be modified towards $T$-dependent effects without major re-definitions of the model assumptions. 
	
	\enlargecolumn{2pt}
	The goal of this work is to establish a $T$-dependent implicit-solvent CG model for PEG, which shall be applicable and transferable in a wide temperature range, even including the LCST, and respects both microscopic structural details and macroscopic thermodynamic properties. For this, we introduce a two-step procedure, combining the microscopic bottom-up approach with a mapping to an effective \emph{analytical} Hamiltonian and a subsequent top-down refinement.  In the first step the interaction (pair) potentials in the CG model are based solely on atomistic simulations in explicit solvent and the microscopic degrees of freedom are integrated out using the IBI method for various temperatures. In the crucial intermediate step, to improve handling and warrant the continuous interpolation (i.e., the transferability) on the $T$-scale in further applications, the interaction potential derived at discrete temperatures is conveniently mapped to a smooth analytical expression, maintaining key structural short-range features like the $T$-dependence of the hydration (desolvation) barrier between monomers. This potential is easy to implement and can be used for any temperature in future simulations. In the final top-down closure,  adjustments of the interaction potential are made in order to match properly the polymer size at room temperature. We demonstrate that this single adjustment together with the $T$-dependence originating from the bottom-up approach is sufficient to successfully reproduce other experimentally known key thermodynamic properties of semi-dilute PEG solutions such as the full equation of state (i.e., osmotic pressure) at various chain lengths as well as their cloud point (or collapse) temperature in the bad solvent condition.
	
	\section{Computational Methods and Coarse-Graining Procedure}
	\label{sec:methods}
	
	\begin{figure}[t]
		\includegraphics[width=\figwidth\linewidth]{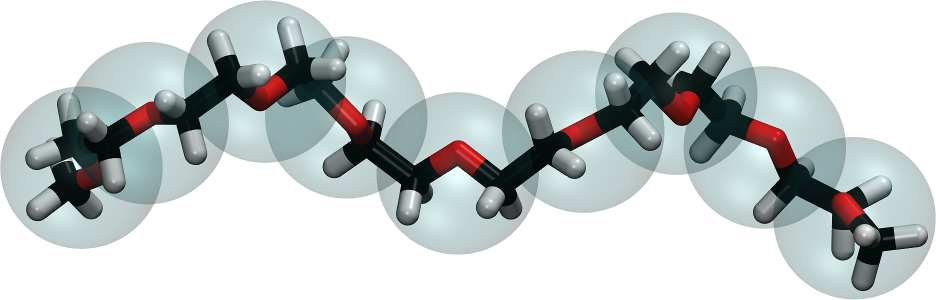}
		\caption{A single PEG nonamer chain, \ce{(EO)_9}, is shown with carbon atoms in black, oxygen in red and hydrogen in white. In the coarse-grained (CG) representation of PEG, the \ce{{}-CH_2-O-CH_2-{}}  group of atoms (or \ce{{}-CH_2-O-CH_3} at the terminal ends) is replaced by a spherical bead (EO unit, depicted as transparent spheres). The center of a CG bead corresponds to the center of the mass of the replaced EO group.}
		
		\label{fig:coarse-grain-unit}
	\end{figure}
	
	\subsection{Atomistic Replica-Exchange Molecular Dynamics (MD) Simulations}
	\label{sec:atomistic-simulation}
	We employed all-atom, explicit-water MD simulations to obtain configurations and interactions of EO oligomers of different lengths (mono-, tri-, and nonamer), cf. the exemplary nonamer in \cref{fig:coarse-grain-unit}.  EO stands for the ethylene oxide unit consisting of \ce{{}-CH_2-O-CH_2-{}} in the main part of the chain or \ce{CH_3-O-CH_2-{}} at the termini, respectively. The free monomer, \ce{CH_3-O-CH_3}, is also described as a single EO unit. The MD simulations were carried out using the Gromacs 4.6 software,\cite{Hess2008,Pronk2013} employing the PEO parametrization developed by Lee \latin{et al.}\cite{Lee2008} with the bonded parameters of the CHARMM force field in combination with the TIP3P water model.\cite{Jorgensen1983} All bonds of the EO oligomers and water molecules were constraint with the LINCS\cite{Hess1997} and SETTLE algorithms,\cite{Miyamoto1992} respectively. The cut-off distance for nonbonded interactions was set to \SI{1.0}{\nano\meter} while long range electrostatics was accounted for by the Particle Mesh Ewald (PME) method with cubic interpolation and a grid spacing of \SI{0.16}{\nano\meter}.\cite{Essmann1995} The long range dispersion correction was applied for energy and pressure.
	
	For the study of $T$-dependent properties of complex liquids and polymers, the replica exchange MD (REMD) method\cite{Swendsen1986} is appropriate to efficiently sample complex energy landscapes in a wide temperature range. We employ the REMD tool as implemented in the Gromacs software.\cite{Hess2008} Here, 48 replicas were generated in a range \SIrange{270}{503}{\kelvin}.\cite{Patriksson2008} Replica exchanges via Monte-Carlo swap moves (accept/reject) were attempted every 50 integration steps. Periodic boundary conditions were used and the individual replicas were simulated under constant pressure and temperature, which were controlled by the velocity-rescale thermostat ($\tau_T = \SI{0.1}{\pico\second}$) and the Parrinello--Rahman barostat (at \SI{1}{\bar}, $\tau_p = \SI{2}{\pico\second}$), respectively.\cite{Bussi2007, Parrinello1981}  Each box contained either 72 monomer and 1984 TIP3P water molecules, or 36 trimer and 1929 TIP3P water molecules, or 36 nonamer and 1404 TIP3P water molecules, yielding approximately \SI{2}{\Molar} (monomer) or \SI{1}{\Molar} (oligomers) concentrations in the water solutions. After the initial energy minimization, the individual replicas were equilibrated in the $NVT$ ensemble for \SI{100}{\pico\second} and in the $NpT$ ensemble for another \SI{100}{\pico\second}. The integration step of the leap-frog (md) integrator was set to \SI{2}{\femto\second} and data were collected every \SI{1}{\pico\second}. The total simulation time per replica was \SI{33}{\nano\second} out of which the last \SI{30}{\nano\second} were used for data analysis.
	
	Recall that the role of the REMD simulation is to provide an accurate (but probably not perfect) atomistic reference for developing the CG potentials in the bottom-up approach, which, together with the iterative Boltzmann inversion (IBI) described further below, is the first step in our CG force field development.
	
	\subsection{Coarse-Grained Simulations}
	\label{sec:coarse-grain-simulation}
	\subsubsection{Computational Methods}
	
	\begin{table}[t!]
		\caption{Parametrization of bonded interactions in the CG model of PEG according to Lee \latin{et al.}\cite{Lee2009}. The bonded interaction energy is defined in \cref{eq:bonded}.}
		\label{tab:bonded}
		\begin{tabular}{|cccc|}
			\hline\Tstrut\Bstrut
			\multirow{2}{*}{bond} & $b^0$ (\si{\nano\meter}) & $k^b$ (\si{\kilo\joule\per\mole\per\nano\meter\squared}) & \\
			& 0.33 & \num{17000} & \\
			\hline\Tstrut\Bstrut
			\multirow{2}{*}{angle} & $\theta^0$ (\si{\deg}) & $k^\theta$ (\si{\kilo\joule\per\mole}) & \\
			& 130 & \num{85} & \\
			\hline\Tstrut\Bstrut
			\multirow{5}{*}{\parbox{\widthof{dihedral}}{\centering dihedral angle}} & $\phi^0$ (\si{deg}) & $k^\phi$ (\si{\kilo\joule\per\mol}) & $n$ \\
			& 180 & 1.96 & 1 \\
			& 0 & 0.18 & 2 \\
			& 0 & 0.33 & 3 \\
			& 0 & 0.12 & 4 \\
			\hline
		\end{tabular}
	\end{table}
	
	In the CG simulations all atoms forming a monomer unit are replaced with a neutral CG bead located in their center of mass, see \cref{fig:coarse-grain-unit}. This symmetric representation (i.e., based on the symmetric \ce{-CH_2-O-CH_2-} group) is reasonable since only small anisotropy effects (e.g., due to a possibly present large molecular dipole) can be neglected.\cite{Fischer2008a, Lee2009, Bedrov2006} The bonded interactions -- which are assumed to be temperature independent -- are based on atomistic simulations as derived in the work of Lee \latin{et al.}\cite{Lee2009} and can be summarized in the interaction energy
	\begin{equation}
	\label{eq:bonded}
	\begin{split}
	U_\text{bonded} 
	&= \sum_\text{bonds} \tfrac12 k^b_{ij} \left(b_{ij} - b^0_{ij}\right)^2\\
	&+  \sum_\text{angles} \tfrac12 k^\theta_{ijk} \left(\cos(\theta_{ijk}) - \cos(\theta^0_{ijk})\right)^2\\
	&+ \sum_\text{torsions} \sum_n  k^\phi_{n,ijkl}\left(1 + \cos(n\phi_{ijkl} - \phi^0_{n,ijkl})\right). 
	\end{split}
	\end{equation}
	The bonded parameters are presented in \cref{tab:bonded}. Only the first neighbors (directly connected by bonds) were excluded from the nonbonded interactions. Furthermore, the water molecules are not present explicitly anymore but implicitly included in the effective $T$-dependent monomer-monomer (pair) interaction potentials. The latter were derived employing the iterative Boltzmann inversion (IBI) approach,\cite{Schommers1983, Reith2003, Rosenberger2016} discussed in detail in \cref{sec:ibi}. 
	
	The CG simulations were performed utilizing the Gromacs 4.6 and 5.1 software,\cite{Hess2008,Pronk2013,Abraham2015} using the stochastic Langevin dynamics (LD) integrator in the $NVT$ ensemble with a friction constant $\gamma = \SI{1}{\per\pico\second}$ and corresponding random force. The mass of the monomer bead is set to \SI{44}{\gram\per\mole} and the integration step could be increased to \SI{10}{\femto\second}. Tabulated potentials with a cut-off distance of \SI{1.2}{\nano\meter} were used during IBI and a cut-off distance \SI{0.9}{\nano\meter} for the final CG potentials. The CG simulations were applied for the same systems as in the atomistic MD simulations to determine the effective CG potentials (bottom-up approach). Therefore the simulation box of a constant volume ($V=\SI{64}{\nano\meter\cubed}$) was filled with 72 monomer, 36 trimer, or 36 nonamer molecules. The simulations were performed for \SI{100}{\nano\second} in a wide temperature range (\SIrange{270}{480}{\kelvin}).
	
	\subsubsection{Calculating PEG solution properties }
	Using the CG simulations, single and many PEG chain properties were calculated, such as the radius of gyration $R_\mathrm{g}$, osmotic pressure or collapse transition (as a marker for the LCST\cite{Wu1998}) as a function of the chain length to relate them to experimental data (top-down approach). Here, single CG polymer chains made of \numlist{9;18;27;36;76;135;275;455;795} EO units were simulated at temperatures \SIlist[list-units=single]{294;320;347;361;371;381;396}{\kelvin}. The analyses, such as regarding the distribution of $R_\mathrm{g}$, were made on trajectories longer than a multiple of the chain correlation (Rouse) time, i.e.,  the shortest chains were sampled for \SI{300}{\nano\second} while the longest chains were sampled for up to \SI{5000}{\nano\second}. The trajectory was split into 20 equally long blocks from which the first two were discarded and the remaining 18 were the subject of the statistical analysis. These simulations were used for the final tuning of the interaction potential, so that the experimental dependence of $R_\mathrm{g}$ and the collapse temperature were reproduced.
	
	For the calculations of the equation of state, 108 relaxed polymer chains, of a length 135, or 455, were placed into the simulation box. LD simulations in the $NpT$ ensemble with a prescribed external pressure (\SIlist{1;10;100;1000}{\kilo\pascal}), controlled by the Parrinello--Rahman barostat with a time constant of $\tau_p = \SI{5}{\pico\second}$, were employed. The total simulation time was \SI{1000}{\nano\second}, employing a time step $\tau = \SI{10}{\femto\second}$, from which the first \SI{100}{\nano\second} were considered as an equilibration phase. The average polymer concentration was determined \latin{a~posteriori} from the mean system volume. It shall be noted that to speed up reaching the desired pressure level, the system was first equilibrated with the Berendsen barostat and an \change{increased} isothermal compressibility of the environment ($\SI{4.5E-3}{\per\bar}$, i.e., \change{one hundred times} the normal isothermal compressibility of water). After the simulation box achieved the presumed volume, the “water-like” isothermal compressibility was restored and the system was equilibrated until no drift in the mean volume was observed.
	
	\subsection{Iterative Boltzmann Inversion}
	\label{sec:ibi}
	
\begin{figure}[t]
	\includegraphics[width=\figwidth\linewidth]{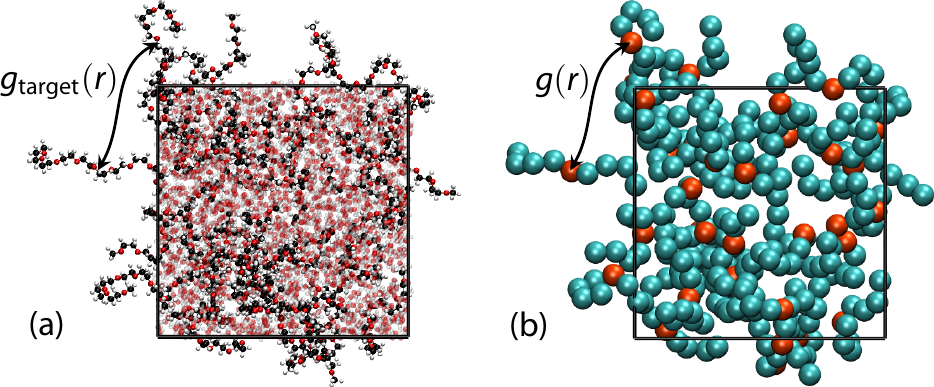}
	\caption{Snapshots of the simulation boxes containing a solution of PEG nonamer as used for the IBI procedure. The molecules only seemingly protrude out of the box as molecules are shown undivided despite the periodic boundary conditions. The pair RDF $g(r)$ between the EO units located right in the middle  of the chains (\nth{5} monomer of the nonamer) in the atomistic simulation (a) is accurately reproduced by the implicit-solvent CG simulation (b). In the CG simulation box, the middle units are depicted in orange while the remaining units are shown as cyan beads.
	\vspace{-2pt}}
	\label{fig:ibi-box}
	\end{figure}

	The iterative Boltzmann inversion (IBI) \cite{Schommers1983, Reith2003, Peter2009, Rosenberger2016} was employed to obtain the effective nonbonded pair potential between the EO units from simulations at finite concentrations of the oligomer. This method is based on iteratively refining the effective potential $U_\text{eff}(r)$ via
	\vspace{-4pt}
	\begin{equation}
	\label{eq:ibi-step}
	%U_{\text{eff}, i+1}(r) = U_{\text{eff}, i}(r) + \lambda RT \ln \frac{g_i(r)}{g_\text{target}(r)}\text,
	U_{\text{eff}, i+1}(r) = U_{\text{eff}, i}(r) + \lambda \change{k_\mathrm{B}}T \ln \frac{g_i(r)}{g_\text{target}(r)}\text,
	\end{equation}
	%\change{where $R$ denotes the gas constant, $T$ the temperature, and $\lambda$ a propagation (damping) constant,}
	\change{where $k_\mathrm{B}$ denotes the Boltzmann constant, $T$ the temperature, and $\lambda$ a propagation (damping) constant,}
	until the structure of the CG solution, expressed by the radial distribution function (RDF) $g_i(r)$, agrees with the target structure from the atomistic simulation $g_\text{target}(r)$. While the coarse-graining procedure works robustly as expected for the solution of simple monomers, faithful mono\-mer-mono\-mer interactions have to respect the connectivity of the chain while avoiding finite chain effects. To overcome this problem, we followed the work of Fischer\cite{Fischer2008a} and performed simulations of solutions of PEG trimers and nonamers, in which the bonded neighborhood is present, yet the chains are not too long, thus still allow for proper sampling. The RDF between the center of mass of the monomer, or the center of mass of the middle EO unit of the oligomer, i.e., the second monomer of the trimer or the fifth monomer of the nonamer (see \cref{fig:ibi-box}), respectively, is then chosen as the reference property.

	\enlargecolumn{8pt}
	The initial effective potential is determined from the Boltzmann inversion of the RDF
	\begin{equation}
	\label{eq:bi}
	%U_{\text{eff}, 0}(r) = - RT \ln g_\text{target}(r)\text{,}
	U_{\text{eff}, 0}(r) = - \change{k_\mathrm{B}}T \ln g_\text{target}(r)\text{,}
	\end{equation}
	respectively from its smoothly decaying variant
	\begin{equation}
	\label{eq:ibi-init}
	%U_{\text{eff}, 0}(r) = - RT \ln g_\text{target}(r) w(r; r_\text{decay})\text{.}
	U_{\text{eff}, 0}(r) = - \change{k_\mathrm{B}}T \ln g_\text{target}(r) w(r; r_\text{decay})\text{.}
	\end{equation}

	The latter is employed, as it is recommended to use only a short range part of such a potential to avoid unwanted oscillations in the long-range part, which may cause convergence problems of the IBI method. This is achieved via application of a smooth weighting function
	\begin{equation}
	\label{eq:ibi-decay}
	w = 
	\begin{cases}
	1 & r \leq r_\text{decay} \\
	\exp(-s(r-r_\text{decay}) & r > r_\text{decay}
	\end{cases}\text{,}
	\end{equation}
	which decays exponentially \change{with a factor $s = \SI{23}{\per\nano\meter}$} beyond a preset distance $r_\text{decay} = \SI{0.95}{\nano\meter}$.
	
	There are known convergence problems of IBI, when initiated with a purely repulsive guess $U_{\text{eff}, 0}(r)$. To overcome this issue for the nonamer, we have employed the initial potential from the trimer instead.
	
	The CG simulations (using $U_{\text{eff}}(r)$) have exactly the same composition as the reference atomistic simulations and were performed at eight different temperatures covering the range from the room to collapse transition temperature. Note that in this work the latter is set equal to the LCST.\cite{Wu1998} The cut-off distance of the nonbonded potential is set to \SI{1.2}{\nano\meter}, which was found sufficient as its increase to \SI{1.5}{\nano\meter} does not influence the resulting effective potentials (\sifigref{3}). 
	
	The solution structure of RDF is rather sensitive to the underlying effective interaction potential and thus a propagation (damping) constant $\lambda$ between 1 and 0.05 was introduced to improve numerical stability of the method.\cite{Bayramoglu2012} As the potential is becoming more complex, a strong damping up to $\lambda = 0.05$ was used in this work. 
	
	\enlargecolumn{8pt}
	The convergence of the IBI method is judged by the evaluation of the merit criteria\cite{Reith2003,Bayramoglu2012}
	\begin{equation} 
	\label{eq:ibi-merit} 
	t_i = \frac{\int {\exp(-r/r_\mathrm{m}) [g_i(r) - g_\text{target}(r)]^2 } {\rm d}r}{\int g_\text{target}(r)^2 {\rm d}r} \text{.}
	\end{equation}
	In this work, we preset the merit criteria threshold to $t < \num{E-5}$ and used $r_\mathrm{m} = \SI{0.41}{\nano\meter}$. Typically up to 30 iteration steps are needed until the merit criteria is reached.

	\section{Reference Experimental Data}
	\label{sec:experimental-data}
	
	\subsection{Collapse transition and LCST behavior}
	\label{sec:experimental-lcst}
	
	The LCST is accompanied by clouding of the solution and is connected to single chain collapse and subsequent aggregation, with both steps driven by increasing strength of hydrophobic interactions. As we are not interested in the fine distinction between these effects, in this work we use LCST, clouding, and collapse transition synonymously.\cite{Wu1998, Grinberg2015, Ashbaugh2006}
	While it is established that the LCST is not present for the short oligomers,\cite{Saeki1976, Bae1991, Saraiva1993} chains up to the 50-mer exhibit a so called closed-loop diagram and possess both a LCST ($\approx \SI{440}{\kelvin}$) and a UCST over a rather broad range of polymer concentrations.\cite{Saeki1976, Bae1991, Saraiva1993} For longer polymer chains the value of the LCST decreases strongly with the polymerization degree up to around 1000-mer, when it saturates around \SI{370}{\kelvin}.\cite{Boucher1976} 
	Interestingly, for the shorter polymer chains the weight fraction also significantly alters the LCST ($\approx\SI{20}{\kelvin}$), while the polydispersity has only a marginal effect.\cite{Saraiva1993} In contrast, LCST of solutions of long polymer chains is much less affected by the polymer concentrations.\cite{Saeki1976, Bae1991, Saraiva1993}  
	
	\subsection{Radius of gyration}
	\label{sec:experimental-gyration}
	
	The radius of gyration $R_\mathrm{g}$ of PEG chain with molecular weight (i.e., polymerization degree \change{$N$}) was determined at around \SI{298}{\kelvin} for a broad range of polymer sizes ($M = \SIrange[retain-unity-mantissa=false,range-units=single]{1E3}{1E7}{\gram\per\mole}$).\cite{Devanand1991,Kawaguchi1997} 
	It was found empirically that the radius of gyration follows the scaling law
	\begin{equation}
	\label{eq:gyration}
	R_\mathrm{g} = lN^\nu\text{,}
	\end{equation}
	with parameters
	$l=\SI{0.1814}{\nm}$ and $\nu=0.58$ \cite{Kawaguchi1997}, or
	$l=\SI[group-digits=false]{0.15195}{\nm}$ and $\nu=0.6$ \cite{Gruijthuijsen2012,Xie2016}.

\change{	
	\subsection{Second virial coefficient}
	\label{sec:experimental-virial}
	The second virial coefficient $B_2$ was experimentally obtained near room temperature employing methods such as light-scattering,\cite{Devanand1991,Kawaguchi1997} freezing point depression,\cite{Wang2002} membrane osmometry\cite{Li2015}, or vapor pressure osmometry\cite{Kushare2013}. The scaling of the second virial coefficient
	\begin{equation}
	\label{eq:virial}
	B_2 = b N^{3\nu}
	\end{equation}
	corresponding to a good solvent regime, with $3\nu \simeq 1.8$, was confirmed by these measurements. The parameter $b$ was determined to be \SI{0.0278}{\nm\cubed}, \cite{Devanand1991} \SI{0.0520}{\nm\cubed},\cite{Wang2002} or \SI[group-digits=false]{0.00583}{\nm\cubed} (with $3\nu=1.81$)\cite{Kawaguchi1997}, respectively. It should be noted that the parameter values span over one order of magnitude.
	
	The temperature dependence of the second virial coefficient was investigated up to \SI{373}{\kelvin} employing small angle X-ray\cite{Pedersen2005} or light\cite{Venohr1998} scattering.
	A linear scaling of $B_2$ with temperature $T$ was proposed  
	\begin{equation}
	\label{eq:virial-temperature}
	%B_2 = B_2^\mathrm{fit} (T_\theta - T)\text{,}
	B_2 = B^\mathrm{fit}(T_\theta - T)\text{,}
	\end{equation}
	with parameters $B^\mathrm{fit} = \SI{2.00}{\nm\cubed\per\kelvin}$ and $T_\theta = \SI{373.2}{\kelvin}$ determined for PEG of \change{molar mass} $M={}$\SI{4600}{\gram\per\mole}\cite{Pedersen2005}.
	
	The conversion of the mass-density based second virial coefficient $A_2$ to the concentration based one $B_2$ is provided in the SI.

%	\begin{equation}
%	\label{eq:virial-a2-temperature}
%	A_2 = A_2^\mathrm{fit} (T_\theta - T)\text{,}
%	\end{equation}
%	with $A_2^\mathrm{fit} = \SI{57E-6}{\per\kelvin}$ and $T_\theta = \SI{373.2}{\kelvin}$.\cite{Pedersen2005}
}

	\subsection{Osmotic pressure}
	\label{sec:experimental-osmotic}
	\enlargecolumn{8pt}
	Recently, osmotic pressure of the PEG polymer samples of the \change{molar mass} $M = {}${\SIlist[list-units=single]{20E3;35E3}{\gram\per\mole}} in concentration up to 2\% (w/w) was measured at \SI{298}{\kelvin} directly by using membrane osmometry.\cite{Li2015} The data were fitted by the Cohen osmotic equation of state (EOS) in the semidilute regime,\cite{Cohen2009}
	\begin{equation}
	\label{eq:cohen}
	\Pi N\,^{\sfrac{9}{5}} = \frac{RT}{M_\mathrm{m} \bar V}\left[\left(\frac{C}{C^*_N}\right) + \alpha \left(\frac{C}{C_N^*}\right)^{\sfrac{9}{4}}\right]\text{.}
	\end{equation}
	Here,  
	%$\Pi$ is the osmotic pressure, $N$ is the number of monomers per chain,
	$M_\mathrm{m}$ is the \change{molar mass} of a monomer ($M_\mathrm{m} = \SI{44.05}{\gram\per\mol}$), $\bar V$ is the partial specific volume of the polymer ($\bar V = \SI{0.825}{\milli\liter\per\gram}$), and $C$ is the polymer mass concentration. The parameter $\alpha$ is the so called crossover index ($\alpha = 0.49$) and $C^*_N$ is the characteristic $N$-dependent polymer concentration. The last two parameters are defined within the Cohen model.\cite{Cohen2009} In particular, $C^*_N \equiv {N^{-\sfrac{4}{5}}}/{\bar V}$, which is a semiquantitative estimation of the polymer concentration up to the semidilute regime.
	
	\change{
	The osmotic pressure of semi-diluted solutions of PEG oligomers (molar mass $M={}$\SIlist{400;1000;4000}{\gram\per\mole}) was measured using vapor pressure osmometry.\cite{Kushare2013} Kirkwood-Buff integrals (KBI)\cite{Kirkwood1951}
	\begin{equation}
		G_{ij} = 4\pi \int_{0}^{\infty} \left(g_{ij}(r)-1\right)r^2\diff{r}\text{,}
	\end{equation}
	where $g_{ij}(r)$ stands for the radial distribution function between particles $i$ and $j$, were calculated for various PEG concentrations from the partial molar volumes.\cite{Kushare2013}
	}
	
	\section{Results and Discussion}
	\label{sec:results}
	
	The CG potential is derived in two steps combining bottom-up and top-down approaches. The first step provides the fine structure of a CG interaction potential which originates mostly from averaging out the solvent degrees of freedom of the explicit-solvent present in atomistic MD simulations via application of the IBI procedure. The interaction potential is then mapped onto an analytical expression. In the second step, the “top-down” closure, the final adjustment of the interaction potential is made in order to match the proper scaling of the  gyration radius with degree of polymerization or collapse transition temperature (LCST). This is later demonstrated to be sufficient to provide a final $T$-dependent model with desired macroscopic properties. 
	
	\subsection{Bottom-up coarse-graining}
	\label{sec:bottom-up}
	
\begin{figure}[t]
		\includegraphics{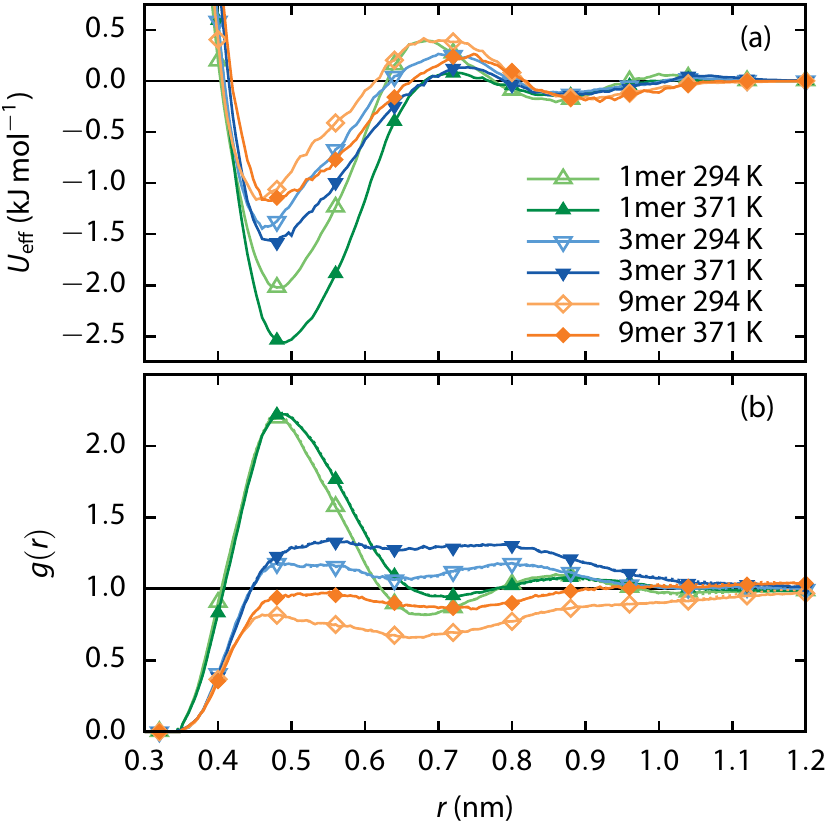}
		\caption[The converged nonbonded pair potentials $U_\text{eff}$ between PEG CG units obtained with IBI using monomer, trimer and nonamer.]
		{(a) The converged nonbonded potentials $U_\text{eff}$ between PEG CG middle beads obtained with IBI using monomers, trimers, or nonamers. (b) Corresponding RDFs  as obtained from atomistic MD simulations and CG LD simulations of molar concentrations (see \cref{fig:ibi-box}) employing the respective converged potentials. An almost perfect match with the corresponding atomistic simulations (dashed lines, mostly overlapped by solid lines of the CG simulations) is achieved.}
		\label{fig:cg-potential}
	\end{figure}

	Selected RDFs between the middle EO units of the simulated oligomers (mono-, tri-, and nonamer) from the atomistic simulations (at finite oligomer concentrations) at ambient temperature and near the expected LCST are shown in \crefx{fig:cg-potential}{b} with dashed lines. A substantial difference in $g(r)$ is found when the systems of free EO monomers and the longer EO oligomers are compared. While the monomer molecules prefer to be in a direct contact at \SI{294}{\kelvin}, as documented by the first peak of $g(r)$ exceeding two, this peak gets close to unity for the trimer and is only about 0.8 in the case of the nonamer. In all systems the affinity between the EO units increases smoothly with temperature, that is, consistent LCST behavior, as can be seen from the comparison between \SIlist{294;371}{\kelvin} in \crefx{fig:cg-potential}{b}.  
	
	\enlargecolumn{2pt}
	To find the effective pair interaction potential which reproduces the EO oligomer solution structure in \crefx{fig:cg-potential}{b}, we apply the IBI procedure, described by \cref{eq:ibi-step}.  As can be seen in \crefx{fig:cg-potential}{b} the solution structure in the CG simulations perfectly reproduces the structure in the atomistic simulations. Notably in the RDFs, the near contacts are becoming less probable for the trimer and are even below one in the whole distance range for the nonamer. This effect in $g(r)$ should be expected, due to the locally increasing excluded volume around the middle bead with increasing polymer chain length. This, however, contrasts with the derived effective nonbonded interaction potentials in \crefx{fig:cg-potential}{a} which are always attractive, although, the depth of the first minimum for the nonamer is only a half of that found in the monomer. More importantly, the effective potential does not change much between the trimer and the nonamer, suggesting a good convergence and saturation behavior and thus a reasonable approximation of the effective interaction potential for long polymer chains at ambient and elevated temperatures.
	 
	Our pair potentials derived at \SI{294}{\kelvin} exhibit similar positions and heights of minima and maxima to those of previous work employing IBI on tri- and decamers, albeit derived from a different force field.\cite{Fischer2008a}
	
	\subsection{Top-down post-refinement}
	\label{sec:top-down}

	\begin{figure}[t]
	\includegraphics{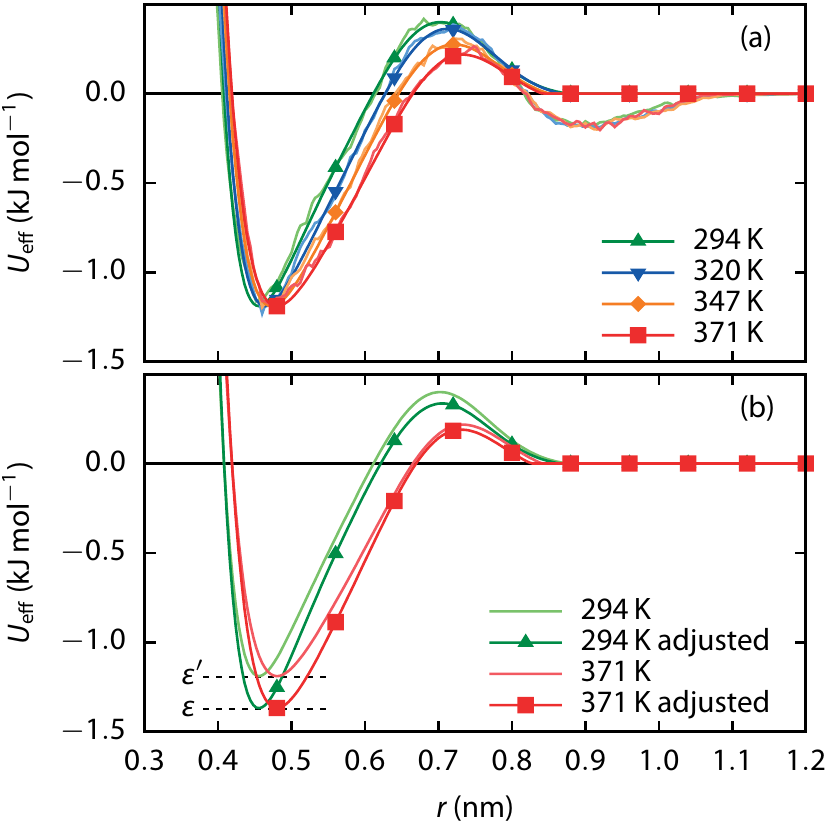}
	\caption[A set of $T$-dependent nonbonded pair potentials between the CG EO units: raw and adjusted.]{
		(a) A set of $T$-dependent nonbonded pair potentials between the CG EO units as obtained by the IBI in a solution of nonamer at various temperatures. Solid lines with symbols are fits by the analytical form \cref{eq:potential} to the IBI raw data (lines without symbols) up to the second maxima. The depth of the first minima is temperature independent, instead the broadening of the potential well causes the greater attraction as the temperature increases. (b) The final adjustment of the effective interaction potentials shown at low (green) and high (red) temperature by shifting the minimum from $\epsilon'$ to $\epsilon$. Fit parameters for the analytical form of the pair potential \cref{eq:potential} are summarized in \cref{tab:fig05-final}.
	}
	\label{fig:potential-fit}
	\end{figure}

	Using the CG potential as directly obtained from the IBI procedure of the nonamer, we determined the mean radius of gyration of long and short PEG chains (\numrange{36}{795}-mer) and compare with the empirical relation of \cref{eq:gyration}.  The CG model, at this stage, systematically underestimates the polymer size for all chain lengths (cf. \sifigref{6}).  To achieve the initial goal, i.e., describe experimental properties of  PEG chains in a whole temperature range as faithfully as possible, we need to apply, as expected, a top-down closure for adjusting the CG potentials by benchmarking to available experimental data. For this purpose, the nonbonded potentials obtained from IBI are mapped to a short-ranged analytic form that is convenient to handle for adjustments and future calculations. We use a form similar to some found in the literature which is a sum of a general Mie potential\cite{Carbone2014} and a Gaussian part. The later represents the hydration (desolvation) barrier and so respects the partially hydrophilic nature of the EO group.\cite{Jusufi2011} Thus, we have
	\begin{equation}
	\label{eq:potential}
	\begin{split}
	U_\text{nonbonded}(r) &= \underbrace{\left( \tfrac{n}{n-m}\right) \left( \tfrac{n}{m}\right)^{m/(n-m)} \epsilon \left[ \left(\frac{\sigma}{r} \right)^{n}- \left( \frac{\sigma}{r}\right)^m \right]}_\text{Mie potential}\\
	&+ \underbrace{\gamma \mathrm{e}^{-\left(\frac{r-\mu}{\delta}\right)^2}}_\text{Gaussian potential}\text{,}
	\end{split}
	\end{equation}
	where $n$ and $m$ are the exponents for the repulsive and attractive interaction terms, respectively, $\epsilon$ the depth of the potential well,  and $\sigma$ the distance where the Mie potential has a zero value; $\gamma$, $\mu$ and $\delta$ are parameters for the Gaussian peak height, position and width. Moreover, analogously as in the raw IBI-derived potentials, we have allowed the parameters of the effective potential to be $T$-dependent. 
	By mapping  onto this form and introducing a potential cut-off at a distance beyond the first maximum, where $U_\text{eff}(r\ge r_{\text{cut-off}})=0$, i.e., we are subtracting the small attraction ($\lesssim 0.1 k_\mathrm{B}T$) which originates from the presence of the second minimum.
	
\begin{figure}[t!]
	\includegraphics{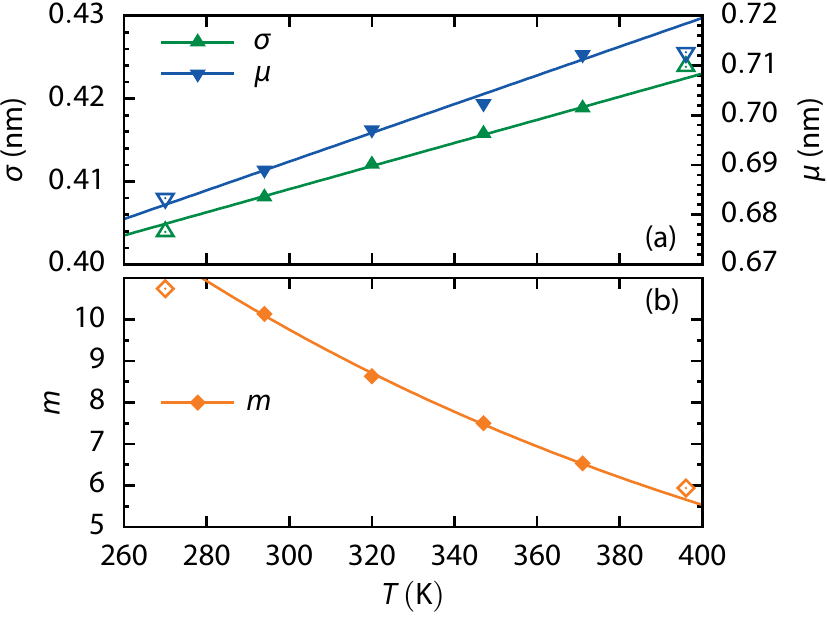}
	\caption[The temperature dependency of potential parameters.]{$T$-dependent parameters of the analytical interaction potential in \cref{eq:potential}. The EO bead size parameters $\sigma$ (green, left axis) and position of the Gaussian potential $\mu$ (blue, right axis) are found to be linearly dependent on $T$ in the approximate range from \SIrange{294}{371}{\kelvin} (LCST):
		$\sigma(T) = (\num{1.39(4)e-4}T/\si{\kelvin} + \num{0.367(1)})\ \si{\nano\meter}$ and $\mu(T) = (\num{2.9(3)e-4}T/\si{\kelvin} + \num{0.604(9)})\ \si{\nano\meter}$.
		The exponent of the attractive  Mie part $m$ can be fitted with the exponential function $m(T) = \num{54(2)} \times \num{0.9943(1)}^{T/\si{\kelvin}}$ in the same $T$-range. Values outside this range are plotted with empty symbols.
	}
	\label{fig:fig05}
	\end{figure}

	The raw IBI potentials can be well fitted by the analytical form at all temperatures, including the second repulsive part, i.e., the desolvation barrier, ($\approx \SI{0.8}{\nano\meter}$), as shown in \crefx{fig:potential-fit}{a}. Interestingly, several fitting parameters are found to be temperature independent, such as $n$, $\epsilon$, $\gamma$ and $\delta$.  The generalized bead size, $\sigma$, and the position of the Gaussian peak, $\mu$, vary linearly with temperature in the range from \SIrange{294}{371}{\kelvin}, see \cref{fig:fig05}. The attractive exponent $m$ shows an exponential dependence in the same temperature range. The set of fitting parameters is summarized in \cref{tab:fig05-final}.

	\begin{table}[t!]
	\caption{The final set of parameters of the CG nonbonded pair potential between EO mers in PEG as described with \cref{eq:potential}. Parameters $m$, $\sigma$ and $\mu$ are functions of temperature $T$. The value of $\epsilon$ is adjusted (from the original fit $\epsilon'$ to the corrected $\epsilon$) in the CG model to correctly reproduce the experimental polymer size scaling with the molecular weight, see text.}
	\label{tab:fig05-final}
	\begin{tabular}{|lll|}
		\hline
		\multicolumn{2}{|c}{parameter} & \multicolumn{1}{l|}{value} \\
		\hline\Tstrut
		Mie repulsion & $n$ & 8.00 \\
		Mie attraction & $m$ & $\num{54} \times \num{0.9943}^{T/\si{\kelvin}}$ \\
		Mie distance & $\sigma$ & $(\num{1.39e-4}T/\si{\kelvin} + \num{0.367})\ \si{\nano\meter}$ \\
		Mie depth & $\epsilon'$ & \SI{1.193}{\kilo\joule\per\mole} \\
		Mie depth corr. & $\epsilon$ & \SI{1.372}{\kilo\joule\per\mole} \\
		Gauss distance & $\mu$ & $(\num{2.9e-4}T/\si{\kelvin} + \num{0.604})\ \si{\nano\meter}$ \\
		Gauss height & $\gamma$ & \SI{0.4841}{\kilo\joule\per\mole}\\
		Gauss width & $\delta$ & \SI{0.1064}{\nano\meter}\\
		\hline
	\end{tabular}
	\end{table}

\begin{figure}[t]
		\includegraphics{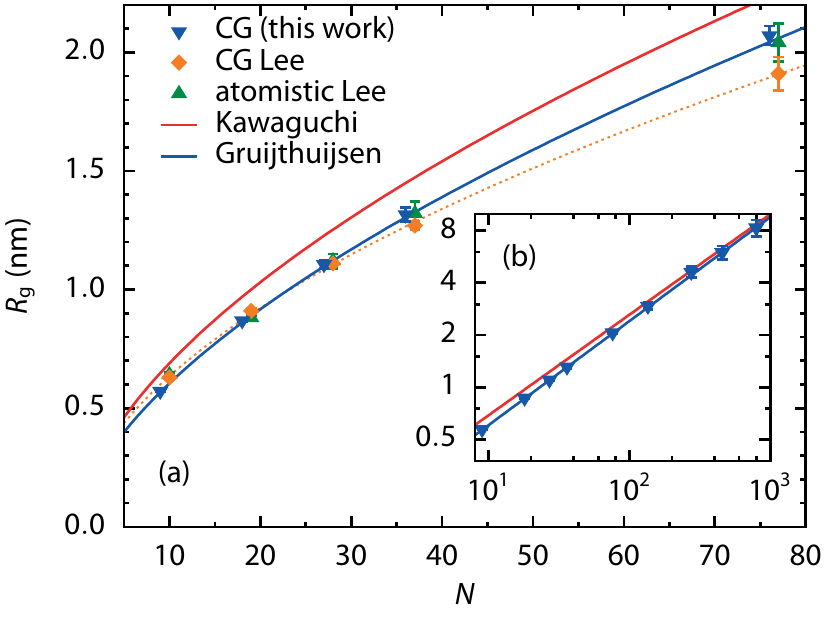}
		\caption[Radius of gyration for PEG chain in water under good solvent conditions obtained from coarse-grained model derived in this work.]{
			Radius of gyration of short (9 to 76-mer) and medium long (up to 795-mer, inset) PEG chains in water under good solvent conditions (\SI{294}{\kelvin}) obtained from the CG model derived in this work (blue triangles). Values are compared with reference atomistic simulations\cite{Lee2008} (green triangles) as well as with a recently proposed CG PEG model with explicit but CG representation of the solvent\cite{Lee2009} (orange squares).
			Finally, we added the extrapolation of the semi-empirical relation determined from experiments with long PEG chains\cite{Gruijthuijsen2012, Xie2016} (blue line). Furthermore, another semi-empirical relation\cite{Kawaguchi1997} is presented for comparison (red line).}
		\label{fig:gyration}
	\end{figure}

	We found that the effective potentials $U_\text{eff}(r)$ originated from the IBI of nonamer resulted in too collapsed states at \SI{294}{\kelvin} when compared to experiments as well as an incorrect distribution of the radius of gyration when confronted with atomistic simulations. This suggests that the IBI-derived pair potential is overall slightly too attractive.  After the mapping onto the analytical form of \cref{eq:potential}, where the second minimum was cut-off, the CG simulation results on the polymer size indicate \change{slightly more} repulsive interactions when compared with the experimental data, as documented in \sifigref{5}. To partly recover the attraction, we empirically adjust only the depth of the first, temperature-independent minima and decrease $\epsilon'$ to the finally adjusted parameter $\epsilon$, also given in \cref{tab:fig05-final}. The exact value of $\epsilon$ was determined by matching the radius of gyration $R_\mathrm{g}$ of 135-mer with the experimental value near room temperature (\SI{294}{\kelvin}). Our final model reaches a very good agreement with experimental data over the whole range of tested PEG chain lengths (36 to 795-mer) as demonstrated in \crefx{fig:gyration}{b}. At the same time the CG model reaches an excellent quantitative agreement with the atomistic simulations for short chains\cite{Lee2008}, see \crefx{fig:gyration}{a}.
	
	Hence, we have presented an analytical form of the $T$-dependent effective CG pair potential between EO units of PEG based on microscopic simulations but fine-tuned to reproduce the experimental data on the $R_\mathrm{g}(N)$ dependence. We emphasize that the original $T$-dependence of three parameters involved in the potential is still present in the model and originates purely from the atomistic simulations, see \cref{tab:fig05-final,fig:potential-fit,fig:fig05}. The $T$-dependent performance and reliability of the model is scrutinized in the next section.

	\subsection{LCST of the optimized CG model}
	\label{sec:results-lcst}
	
\begin{figure}[t!]
	\includegraphics{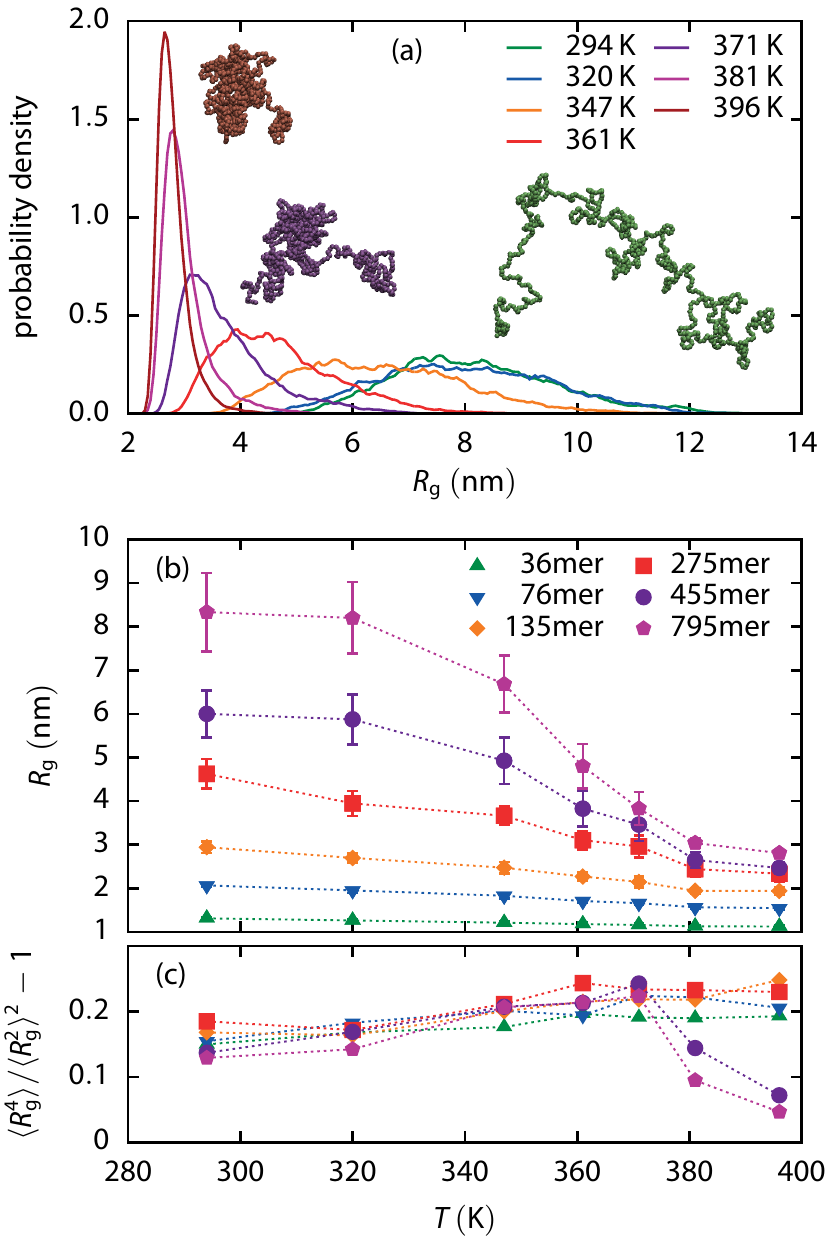}
	\caption[The radius of gyration of PEG in water as a function of temperature for single polymer chain of various lengths determined from LD simulation with proposed implicit-solvent CG model.]{
		(a) Distribution of the gyration radius $R_\mathrm{g}$ of a 795-mer in the temperature range \SIrange{294}{396}{\kelvin}. With increasing temperature, the extended conformations are decreasingly populated, followed by a deformed and broaden distribution around the LCST (\SIlist{361;371}{\kelvin}) to ultimately only exist in highly compact states at the highest temperatures. The inset graphics present the typical conformation at given temperature (i.e., \SIlist{294;371;396}{\kelvin}) with polymer colors matching the colors of the respective data lines. (b) The radius of gyration of aqueous PEG and (c) its fluctuation as a function of $T$ for a single polymer chain of various lengths determined from the CG simulation. The radius of gyration decreases sharply between \SIrange{350}{380}{\kelvin} for 455-mer and 755-mer which agrees well with the experimentally determined value of LCST around \SI{371}{\kelvin}. In chains up to 275-mer ($M = \SI{12100}{\gram\per\mole}$), fluctuations increase monotonically with temperature between \SIrange{294}{396}{\kelvin}. Longer polymer chains, however, exhibit a maximum in the fluctuations around \SI{371}{\kelvin} followed by an abrupt decrease as the chains collapse to the globule. We define the collapse transition temperature or LCST as the temperature with maximal fluctuations.}
	\label{fig:gyration-temperature}
	\end{figure}

	PEG solutions become cloudy around a temperature of \SI{370}{\kelvin} where water turns to  a bad solvent of PEG and is often identified as the LCST. The clouding is due to chain collapse and subsequent aggregation, with both steps driven by increasing strength of hydrophobic interactions. 
	We perform a series of CG simulations of PEG chains consisting of 36 to 795 mers in the temperature range from \SIrange{294}{396}{\kelvin} to verify the $T$-dependent collapse behavior of our optimized CG model.  The CG simulations in the range \SIrange{294}{381}{\kelvin} were performed using a parametric interpolation, while outside this range, i.e., at \SI{396}{\kelvin}, parameters from \sitabref{1} were directly used. In the first step of the analysis, the mean radius of gyration of the PEG chains is calculated in this temperature range, cf. \cref{fig:gyration-temperature}, and the existence and location of the collapse transition is compared to the known experimental data.\cite{Saeki1976}
	
	While for the shorter chains ($<$135-mer) the collapse is rather \change{gradual} in the whole investigated temperature range, for the long chains of 455-mer and 755-mer, the mean radius of gyration has a sigmoidal shape and decreases rather sharply around \SIrange{350}{380}{\kelvin}, see \crefx{fig:gyration-temperature}{b}. We then define the collapse transition (or LCST) as the temperature where the single chain collapses and, typical for a critical transition of a finite system, the radius of gyration maximizes its fluctuations. The suggested measure of these fluctuations is the quantity $\langle R_\mathrm{g}^4 \rangle / \langle R_\mathrm{g}^2 \rangle^2 - 1$.\cite{Ivanov1998} We thus focus on the distribution of the radius of gyration as a function of temperature and quantify the conformational fluctuations of the chain, see \crefx{fig:gyration-temperature}{c} and \sifigref{10}.

	For shorter chains, up to the 275-mer, the fluctuations increase monotonically with temperature between \SIrange{294}{396}{\kelvin}, while for longer polymer chains, maximum fluctuations around \SI{371}{\kelvin} followed by an abrupt decrease of fluctuations in the collapsed globular state are observed in \crefx{fig:gyration-temperature}{c}. Based on this analysis, we determine a LCST $\approx\SI{371}{\kelvin}$ for longer chains, while no critical temperature (LCST or UCST) is observed for short chains within our CG model and analysis. These findings are consistent with experimental data, where low concentration solutions of low mass PEG have the cloud point shifted to significantly higher temperatures ($>\SI{400}{\kelvin}$) or no critical point is observed at all (for chains shorter than the 52-mer).\cite{Saeki1976}
	
	Based on the above stated analysis, we conclude that the derived CG model with $T$-dependent parameters is able to reproduce the LCST behavior without further tweaking of the parameters. Our model is thus proved to be able to reproduce and describe single chain properties (i.e., at infinite dilution of the polymer) in a broad range of temperatures and chain lengths. In the next section, we will apply the CG model also at finite polymer concentrations to scrutinize its behavior and applicability for more complex systems.
	
	\subsection{Osmotic pressure of the optimized CG model}
	\label{sec:results-osmotic}

\begin{figure}[t]
		\includegraphics{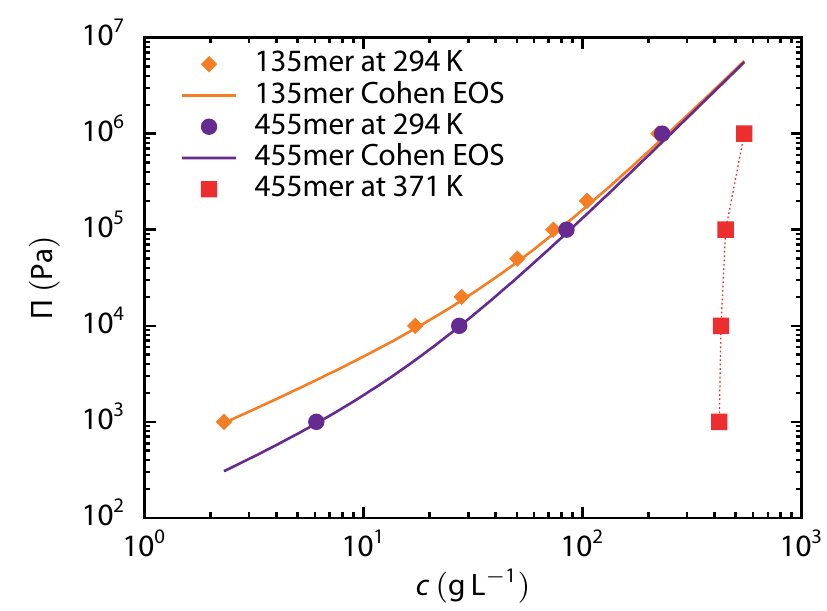}
		\caption[Osmotic pressure]{
			The osmotic pressure $\Pi$ of PEG as a function of the polymer mass concentration in water for various chain lengths at \SI{294}{\kelvin} and \SI{371}{\kelvin}. The CG simulation data, shown as symbols (error bars are of symbol size), are compared with the Cohen EOS (\cref{eq:cohen}) valid for the semi-dilute regime far below the LCST (solid lines). The thin dotted line for the \SI{371}{\kelvin} simulation data is just a guide to the eye.}
		\label{fig:osmotic-pressure}
	\end{figure}

	In order to compare to the experimental osmotic pressure, as described in \cref{sec:methods}, we perform CG simulations in the $NpT$ ensemble of systems containing 108 chains of either 135-mers or 455-mers.  All the simulations are performed at ambient temperature, where experimental data are available, as well as close to the LCST, where interesting physics is expected, due to the polymer collapse and the strong temperature dependence of the inter-chain interactions. Results are presented in \cref{fig:osmotic-pressure}.  Data obtained from the simulations in a good solvent regime at \SI{294}{\kelvin} are compared with the Cohen EOS, \cref{eq:cohen}, valid for a semidilute concentrations with experimentally determined parameters $\bar V$ and $\alpha$ at \SI{298}{\kelvin}. In the investigated concentration ranges up to $\approx 20$\% by weight, the osmotic pressure obtained by simulations agrees well with the values predicted.  At the LCST, the concentration of PEG is high ($\approx\SI{400}{\gram\per\liter}$) even for the lowest simulated osmotic pressure \SI{1}{\kilo\pascal} as expected for the more concentrated phase in the phase separated regime. At the LCST the Cohen EOS is not valid and no fitting was attempted.

	We finally note that in the low density / low pressure limit of our data at \SI{1}{\kilo\pascal} \change{and \SI{294}{\kelvin}} the weight concentration of PEG is 0.23\% for the 135-mer and 0.61\% for the 455-mer, respectively, see \cref{fig:osmotic-pressure}. \change{Applying the leading order viral expansion, $\Pi = k_\mathrm{B}T(c+B_2 c^2)$, to the smallest concentration, gives the values for the second virial coefficient, $B_2^\mathrm{EOS} = {}$\SIlist{240;1890}{\nano\meter\cubed} for the 135-mer and 455-mer, respectively. Evaluation of the KBI in the same low density limit provides the second virial coefficient $B_2 = - G_{22}/2$, yielding $B_2^\mathrm{KBI}={}$\SIlist{150;1100}{\nano\meter\cubed} for the 135-mer and 455-mer, respectively.} Experimental data\cite{Devanand1991} are \SIlist{190;1300}{\nano\meter\cubed} for the 135-mer and 455-mer, respectively, reasonably close to our predictions. Hence, we have demonstrated that our model reliably reproduces the EOS of aqueous PEG for a wide range of temperatures, molecular weights, and concentrations.

	\change{
	\subsection{Kirkwood-Buff Integrals}
	The Kirkwood-Buff theory is a rigorous approach for the structure-thermodynamics relation of solutions and mixtures of arbitrary number of components, in which the KBIs contain full information about the solution structure at a given system composition. In contrast to the virial coefficients, the KBIs are concentration and temperature dependent functions, and their combinations directly provide macroscopic properties, such as PEG activity coefficients and partial molar volumes.

	\begin{figure}[t]
		\includegraphics{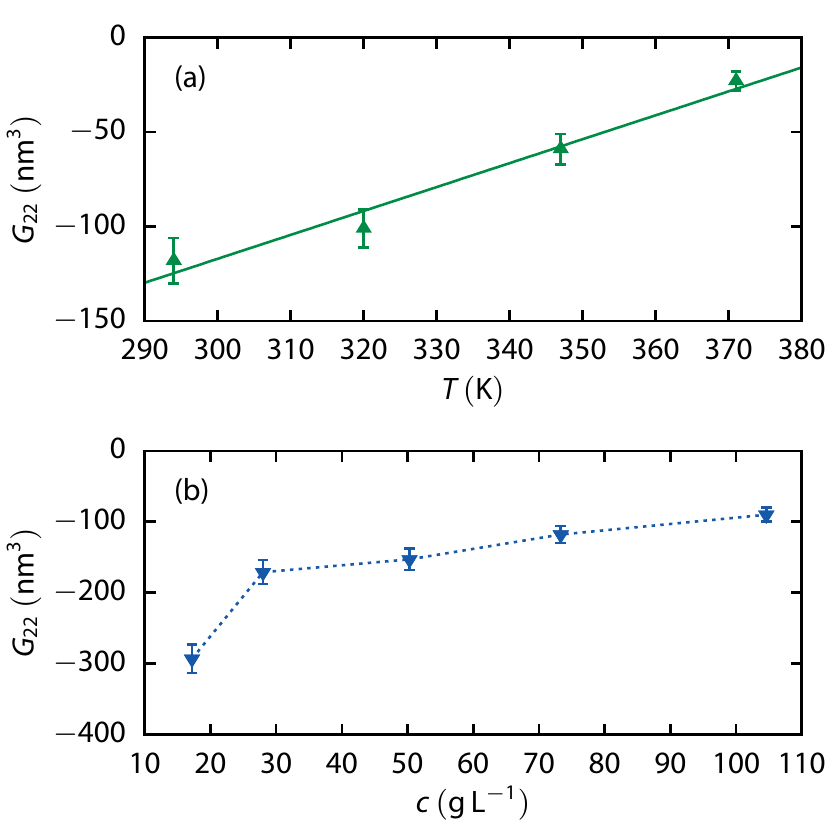}
		\caption[Kirkwood-Buff integrals]{\change{The Kirkwood-Buff integrals $G_{22}$ (symbols) between PEG 135-mer chains. (a) Under the constant osmotic pressure \SI{1}{\bar}, the KBI is evaluated as a function of temperature. The linear regression of the data is presented with a solid green line. (b) The concentration dependence of the KBI in a good solvent regime at \SI{294}{\kelvin}. The thin dotted blue line is a guide to the eye.}}
		\label{fig:kirkwood-buff}
	\end{figure}
	
	In the coarse-grained description of a PEG solution, the only remaining KBI is $G_{22}$ between polymer chains, which temperature dependence is presented in \crefx{fig:kirkwood-buff}{a}. $G_{22}$ is negative, but increases with temperature, which indicates that the “aggregation” forces strengthen with temperature. This is consistent with the measured temperature dependence of the second virial coefficient,\cite{Venohr1998, Pedersen2005} and culminates as LCST behavior at elevated temperatures, where the PEG conformation entropy and other polymer-water interaction effects must be also taken into account.
	
	In \crefx{fig:kirkwood-buff}b we can see a positive slope of $G_{22}$ with concentration, which reflects the strengthening of the polymer-polymer affinity with increasing amount of interacting polymers. These findings agree with the concentration dependence of $G_{22}$ determined experimentally.\cite{Kushare2013}
	}

	\section{Conclusion}
	\label{sec:conclusions}
	
	Using a combined bottom-up and top-down approach a fully implicit-solvent CG model of PEG was developed that is transferable between different temperatures (that is, different solvent qualities). In the first step, the IBI procedure was employed on the solution of EO oligomers to integrate out the solvent and polymer atomistic degrees of freedom and thus determine the fine structure of the effective nonbonded pair potentials between the EO units. The solutions of nonamer were selected for this purpose as they sufficiently include the effects of bonded neighbors and the influence of terminal groups is suppressed. The coarse-graining procedure was employed in a wide temperature range (\SIrange{270}{450}{\kelvin}) and an analytic form of the nonbonded potential with three $T$-dependent parameters was proposed. In a crucial intermediate step, the atomistic structure up to the desolvation barrier was faithfully retained in the analytic expression for simple and continuous transferability. In the top-down closure, a temperature independent parameter was adjusted to achieve agreement with the experimentally known radius of gyration of the 135-mer.
	
	The post-refined model was tested under good solvent conditions for polymer size and pressure as well as for LCST prediction. At \SI{294}{\kelvin} excellent agreement of the radius of gyration as a function of polymer chain length was achieved comparing  both to the atomistic simulation (for short polymer chains) and the available experimental data (up to 795-mer $\sim \SI{35000}{\gram\per\mole}$). At finite concentrations, the osmotic pressure was determined in direct simulation and compared with experimental data. The Cohen equation of state for semidilute concentrations was obeyed by the model. Furthermore the second virial coefficient $B_2$ was estimated to be in a good agreement with experimentally determined values. Finally, for a sufficiently long single chains the peak in $R_\mathrm{g}$-fluctuations and the coil-to-globule transition at elevated temperature (\SI{370}{\kelvin}) under bad solvent conditions is observed. This clear signature of criticality compares well to the experimentally determined value of the cloud point temperature. Finally, at finite concentrations, an initially diluted system having low osmotic pressure becomes highly concentrated when the temperature is elevated to \SI{370}{\kelvin}, resembling the dense phase expected above the LCST.
	
	Our transferable CG model should be useful for efficient future simulations of aqueous PEG in more complex systems at various temperatures and close to the important collapse (switching) transition, pushing coarse-grained simulations further into a potential applicability to soft functional material design.

	\begin{acknowledgement}
		\label{sec:acknowledgement}
		R.\,C. and J.\,D. acknowledge funding from the Deutsche Forschungsgemeinschaft (DFG grant DZ-74/6), Germany, for this project.  J.\,H. thanks the Czech Science Foundation (grant 16-57654Y) for support.
	\end{acknowledgement}

	\begin{suppinfo}
		 MARTINI force field evaluation; IBI results for monomers and trimers; values of the nonbonded potential parameters (data in \cref{fig:fig05}); details of the top-down refinement; structural comparison between the atomistic and CG model; simulation trajectories of the radius of gyration; KBI for short oligomers.
	\end{suppinfo}

\providecommand{\latin}[1]{#1}
\makeatletter
\providecommand{\doi}
  {\begingroup\let\do\@makeother\dospecials
  \catcode`\{=1 \catcode`\}=2\doi@aux}
\providecommand{\doi@aux}[1]{\endgroup\texttt{#1}}
\makeatother
\providecommand*\mcitethebibliography{\thebibliography}
\csname @ifundefined\endcsname{endmcitethebibliography}
  {\let\endmcitethebibliography\endthebibliography}{}

\end{document}